\documentclass[final,3p,times]{elsarticle}

\usepackage{amssymb}
\usepackage{amsmath}
\usepackage{multicol}
\usepackage{wrapfig}
\usepackage{needspace}
\usepackage[colorlinks,allcolors=blue]{hyperref}

\journal{Journal of Subatomic Particles and Cosmology}

\begin{document}

\begin{frontmatter}

\title{Hadron resonance gas with density-dependent interactions
       for neutron stars and heavy-ion collisions}

\author[uh]{Volodymyr Vovchenko}
% \ead{vvovchenko@uh.edu}
\author[uh]{Volodymyr Kuznietsov}
\author[uh]{Tripp Moss}

% \cortext[cor]{Corresponding author.}
\affiliation[uh]{organization={Department of Physics, University of Houston},
             addressline={3507 Cullen Blvd},
             city={Houston},
             postcode={77204},
             state={TX},
             country={USA}}

\begin{abstract}
We present a density-dependent generalization of the van der Waals hadron
resonance gas model (DD-HRG) for describing both the hot hadronic matter
created in heavy-ion collisions and the cold, dense matter inside neutron
stars. 
Non-resonant interactions are incorporated through a generalized excluded-volume prescription with a density-dependent available-volume fraction, supplemented by an arbitrary density-dependent mean field.
With isospin-dependent interaction parameters constrained by empirical properties of nuclear matter, the resulting equation of state extends the causality range to include neutron-star interiors and supports two-solar-mass stars. 
It also improves the description of lattice QCD thermodynamics and conserved-charge susceptibilities at vanishing baryochemical potential, with lattice data favoring reduced repulsion among strange baryons.
This DD-HRG framework is available within the latest version of the open-source \texttt{Thermal-FIST} package.
\end{abstract}


\end{frontmatter}

\section{Introduction}
\label{sec:intro}

The hadron resonance gas (HRG) model describes strongly interacting matter below the QCD pseudocritical temperature as a multi-component gas of the known hadrons and resonances. 
Its theoretical basis is the relativistic virial (S-matrix) expansion, in which attractive hadronic interactions are dominated by resonance formation~\cite{Dashen:1969ep,Venugopalan:1992hy}. 
The ideal HRG model reproduces lattice QCD thermodynamics at $\mu_B = 0$ for $T \lesssim T_{\rm pc}$ and provides a thermal model description of hadron
abundances in heavy-ion collisions, 
with many extensions implemented in open-source packages such as \texttt{Thermal-FIST}~\cite{Vovchenko:2019pjl}.

Quantitative applications are, however, sensitive to non-resonant
interactions, which have been incorporated through excluded-volume (EV)
corrections~\cite{Yen:1997rv,Vovchenko:2017xad}, repulsive mean
fields~\cite{Huovinen:2017ogf}, the S-matrix
formalism~\cite{Andronic:2018qqt}, and effective-mass
approaches~\cite{Savchuk:2020yxc,Vovchenko:2020crk}, with parameters constrained by, e.g.,
lattice susceptibilities~\cite{Karthein:2021cmb} or proton number
cumulants~\cite{Vovchenko:2021kxx}. 
The van der Waals (VDW) HRG~\cite{Vovchenko:2016rkn} combines  excluded-volume repulsion with an intermediate-range attraction and thereby captures the nuclear liquid--gas transition, whose remnants influence fluctuation observables
even at $\mu_B = 0$~\cite{Mukherjee:2016nhb}. 
Since the liquid--gas transition is driven by nuclear interactions, 
the interacting HRG extends in principle to the
neutron-star (NS) corner of the QCD phase diagram. 
An HRG model with a Carnahan--Starling refinement of the EV repulsion and constant VDW parameters indeed yields
gravitationally bound neutron stars~\cite{Fujimoto:2021dvn}.

Two obstacles have so far prevented a quantitative HRG description of NS
matter: firstly, the standard VDW excluded volume makes the $T=0$ equation of
state (EoS) excessively stiff with incompressibility $K_0 \simeq 760$~MeV~\cite{Vovchenko:2017cbu},
and enforces a relatively low hard-packing limit on baryon density at $n_B \sim 1.8\,n_0$. 
Secondly, the extended formulations are limited to single-component interactions~\cite{Vovchenko:2017cbu}.
As such, they are isospin-blind and yield partly acausal NS branches~\cite{Fujimoto:2021dvn}. 
In this contribution, which builds on
Refs.~\cite{Vovchenko:2017cbu,Moss:2024zwd} and the implementation within
\texttt{Thermal-FIST}~\cite{Vovchenko:2019pjl}, we present a
\textit{density-dependent} HRG (DD-HRG) that mitigates both obstacles,
providing a causal, isospin-dependent EoS that supports
$M \sim 2\,M_\odot$ neutron stars while remaining consistent with
lattice QCD constraints at $\mu_B = 0$.

\section{Density-dependent van der Waals HRG}
\label{sec:formalism}

\subsection{Single-component formulation}

The DD-HRG generalizes both the excluded-volume and mean-field
prescriptions to arbitrary functions of density. For a single component,
the Helmholtz free energy reads~\cite{Vovchenko:2017cbu}
\begin{equation}
F(T,V,N) = F_{\rm id} \big(T,\,V f(\eta),\,N\big) + V\,v(n),
\label{eq:F1}
\end{equation}
where $n = N/V$, $f(\eta) \leq 1$ is the available-volume fraction written
in terms of the packing fraction $\eta = bn/4$, and $v(n)$ is a
density-dependent mean-field energy density. 
Familiar choices of $f$ are~\cite{Moss:2024zwd}
\begin{equation}
f_{\rm VDW} = 1-4\eta, \qquad
f_{\rm CS} = e^{-\frac{(4-3\eta)\eta}{(1-\eta)^{2}}}, \qquad
f_{\rm TVM} = e^{-4\eta-8\eta^{2}}.
\label{eq:fchoices}
\end{equation}
The VDW form corresponds to $V \to V - bN$ and reaches hard packing at
$n = 1/b$. 
The Carnahan--Starling (CS) form extends the packing limit fourfold while the
tri-virial model (TVM) reproduces the same low-density virial coefficients but removes the packing singularity altogether~\cite{Vovchenko:2019hbc}. 
The excluded volume
suppresses single-particle occupation numbers,
$\rho(p) \to f(\eta)\,\rho(p)$, leaving the Fermi sea unsaturated at $T=0$, which effectively mimics the quark Pauli exclusion principle, in
analogy with quarkyonic
matter~\cite{Poberezhniuk:2023rct,Fujimoto:2023unl}. 
The mean field typically models attractive interactions, although it can also describe repulsion, e.g., in the case of the Skyrme mean field.
Equivalently, the construction
amounts to density-dependent VDW parameters,
$a_{\rm eff}(n) = -v(n)/n^{2}$ and $b_{\rm eff}(n) = [1-f(\eta)]/n$.
All other thermodynamic quantities follow from $F(T,V,N)$ via standard thermodynamic relations.

\subsection{Multi-component generalization and Thermal-FIST}

For a mixture of hadron species $i$, the free energy generalizes to
\begin{equation}
F(T,V,\{N_i\}) = \sum_{i} F_{i}^{\rm id} \big(T,\,V f_i(\{n\}),\,N_i\big) + V\,v(\{n\}),
\label{eq:Fmulti}
\end{equation}
where both the available-volume fractions $f_i(\{n\})$ and the mean field
$v(\{n\})$ may depend on all hadron densities $\{n\}$. The thermodynamics
follows from a closed system of transcendental equations for the effective
chemical potentials $\mu_i^*$,
\begin{equation}
\mu_i = \mu_i^* - \sum_k f_{k,i}\, p_k^{\rm id}(T,\mu_k^*) + v_i,
\qquad
n_i = f_i\, n_i^{\rm id}(T,\mu_i^*),
\label{eq:trans}
\end{equation}
with $f_{k,i} \equiv \partial f_k/\partial n_i$ and
$v_i \equiv \partial v/\partial n_i$, where $p^{\rm id}$ and $n^{\rm id}$
are ideal-gas functions. Special cases include the
EV-HRG~\cite{Yen:1997rv}, VDW-HRG~\cite{Vovchenko:2016rkn}, and
mean-field HRG ($f_i = 1$)~\cite{Huovinen:2017ogf}.

The DD-HRG thermodynamics is fully implemented in the open-source
\texttt{Thermal-FIST} package~\cite{Vovchenko:2019pjl} since version 1.5,
with arbitrary $v(\{n\})$ and $f_i(\{n\})$ specified through C++ classes
in its ``real-gas'' module. 
Other extensions introduced in versions 1.5--1.6 include leptons and $\beta$-equilibrium, neutron-star and cosmic trajectories, magnetic fields, the pressure Hessian, the PDG2025-based particle list, an updated charm list, and a browser-based WebAssembly interface~\cite{ThermalFIST:WASM}.

\begin{figure}[t]
\centering
\includegraphics[height=5.2cm]{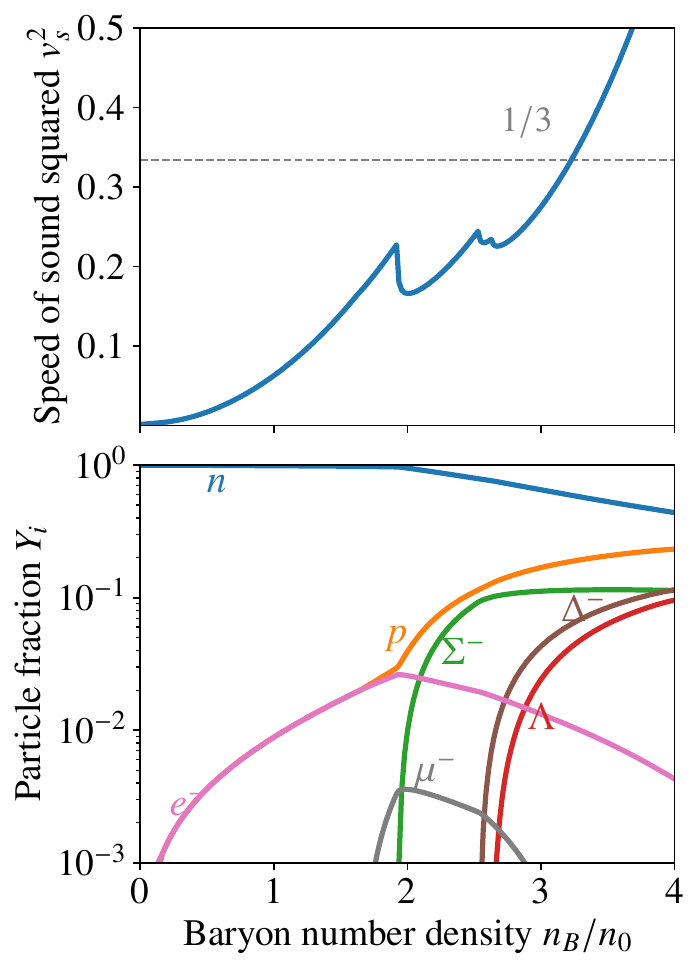}\hspace{7mm}%
\includegraphics[height=5.2cm]{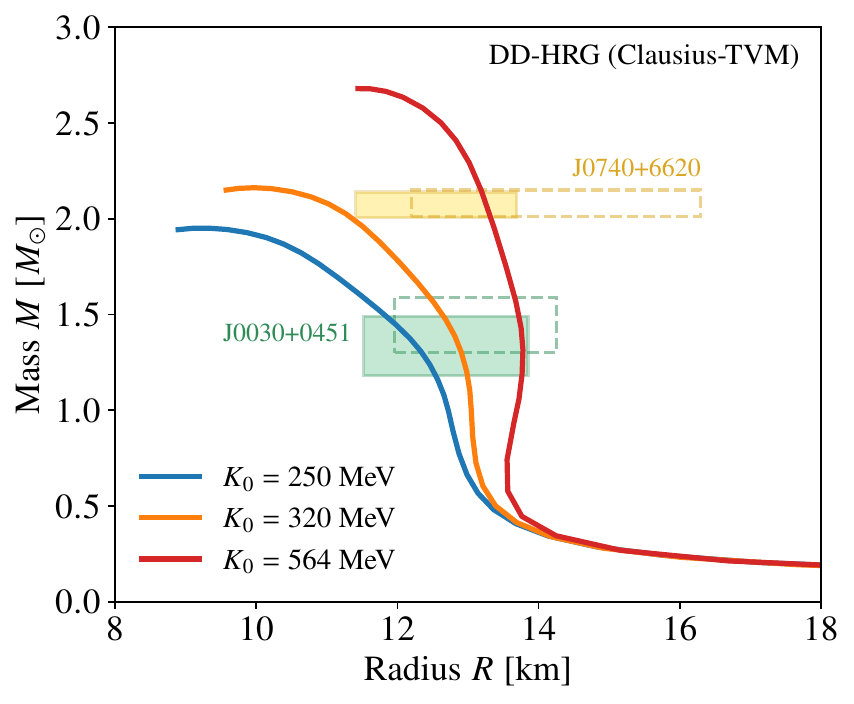}
\caption{Left: speed of sound squared (top) and particle fractions
(bottom) in $\beta$-equilibrated neutron-star matter in the DD-HRG
(Clausius--TVM) model with $K_0 = 320$~MeV. Right: the corresponding
mass--radius relations for $K_0 = 250$, $320$, and $564$~MeV, the latter
corresponding to the pure VDW mean field ($c = 0$). The rectangles depict
the NICER mass--radius estimates for PSR~J0030$+$0451 and PSR~J0740$+$6620
(filled: Riley et al.~\cite{NICER}, dashed: Miller et al.~\cite{NICER2}). }
\label{fig:ns}
\end{figure}

\section{Neutron-star matter}
\label{sec:ns}

In a baryonic mixture of nucleons, hyperons, and resonances at charge
fraction $y = \rho_Q/\rho_B$, the pair interaction parameters
depend on the quantum numbers of the interacting baryons.
Here we distinguish
isospin-like ($a_{nn}$, $b_{nn}$) and isospin-unlike
($a_{pn}$, $b_{pn}$) channels. 
Symmetric nuclear matter corresponds to
$a=(a_{nn}+a_{pn})/2$ and $b=(b_{nn}+b_{pn})/2$.
Following~\cite{Moss:2024zwd}, we combine a Clausius-like mean field with the TVM available volume,
\begin{equation}
v(\{n\}) = -\sum_{i,j\in B}\frac{a_{ij}\,n_i n_j}{1+c\,n_B}, \qquad
f_i(\{n\}) = e^{-4\eta_i-8\eta_i^{2}}, \qquad
\eta_i = \frac14\sum_{j\in B} b_{ij}\,n_j,
\label{eq:clausius-tvm}
\end{equation}
applied to all baryon pairs, with the analogous (separate) construction in
the antibaryon sector. Note that the Clausius-like mean field smoothly converges to VDW attraction as $c \to 0$. The binding energy $E/A = -16$~MeV at saturation
density $n_0 = 0.16$~fm$^{-3}$ fixes $a \approx 350$~MeV~fm$^3$ and
$b \approx 4.3$~fm$^3$ for the pure VDW mean field ($c=0$) combined with TVM repulsion, which,
however, yields a stiff EoS with incompressibility $K_0 = 564$~MeV. 
The
Clausius denominator softens the EoS: the roughly empirical range
$K_0 \approx 250$--$320$~MeV is recovered for
$c \approx 4.5$--$3.0$~fm$^3$, with $a$ and $b$ refitted
accordingly to preserve ground-state properties~\cite{Moss:2024zwd,Lysenko:2024ulc}. The empirical symmetry
energy ($J \approx 30$--$35$~MeV) and its slope ($L \approx 59$~MeV) are used to fix
the isospin splitting, $a_{pn}/a_{nn} \approx 2.5$--$2.7$ and
$b_{pn}/b_{nn} \approx 1.6$--$1.8$. 
This implies considerably stronger
interactions in the isospin-unlike channel.

Neutron-star matter is obtained by adding electrons and muons as ideal
Fermi gases and imposing electric charge neutrality and
$\beta$-equilibrium, $\mu_n - \mu_p = \mu_e = \mu_\mu$, while strangeness is assumed to be in equilibrium, $\mu_S = 0$.
Our preliminary results for the composition and speed of sound at $T=0$ are
shown in Fig.~\ref{fig:ns} (left) for $K_0 = 320$~MeV. Heavy baryons
appear sequentially, $\Sigma^-$ at $n_B \simeq 2\,n_0$, $\Delta^-$ and
$\Lambda$ at $n_B \simeq 2.6$--$2.7\,n_0$, with each onset
producing a kink and temporary softening in $v_s^2$. 
The EoS remains
causal throughout the densities probed, with $v_s^2$ exceeding $1/3$ only above $n_B \simeq 3.2\,n_0$. 
The Tolman--Oppenheimer--Volkoff
mass--radius relations are shown in Fig.~\ref{fig:ns} (right): for
$K_0 = 250$--$320$~MeV the model
supports maximum masses $M_{\rm max} \simeq 1.95$--$2.15\,M_\odot$,
in line with two-solar-mass pulsars and NICER radius
measurements~\cite{NICER,NICER2}, while the stiff VDW
limit reaches $M_{\rm max} \simeq 2.7\,M_\odot$.

\begin{figure}[t]
\centering
\includegraphics[width=0.365\linewidth]{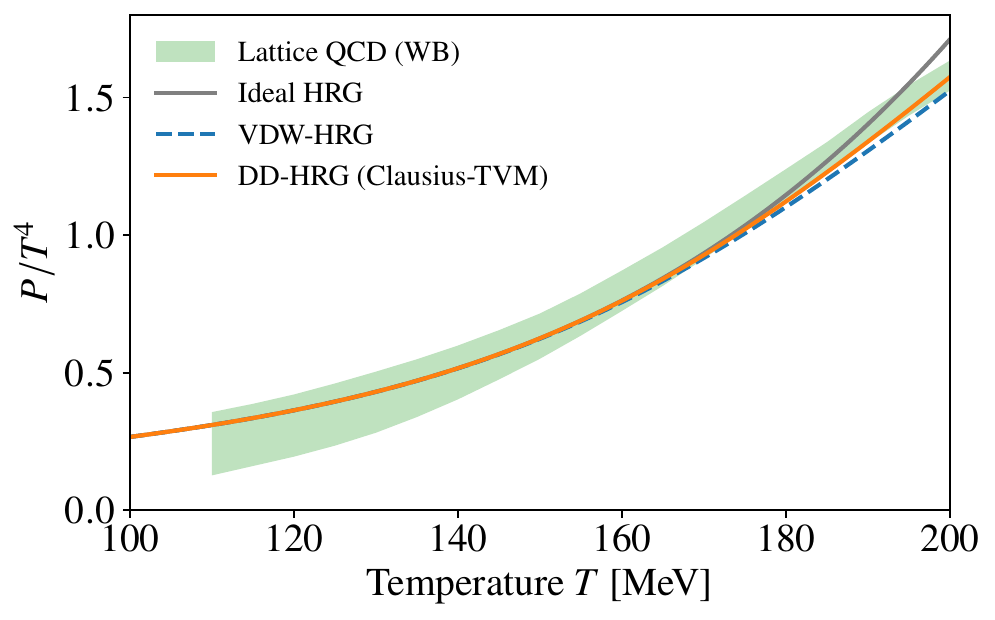}\hspace{5mm}%
\includegraphics[width=0.365\linewidth]{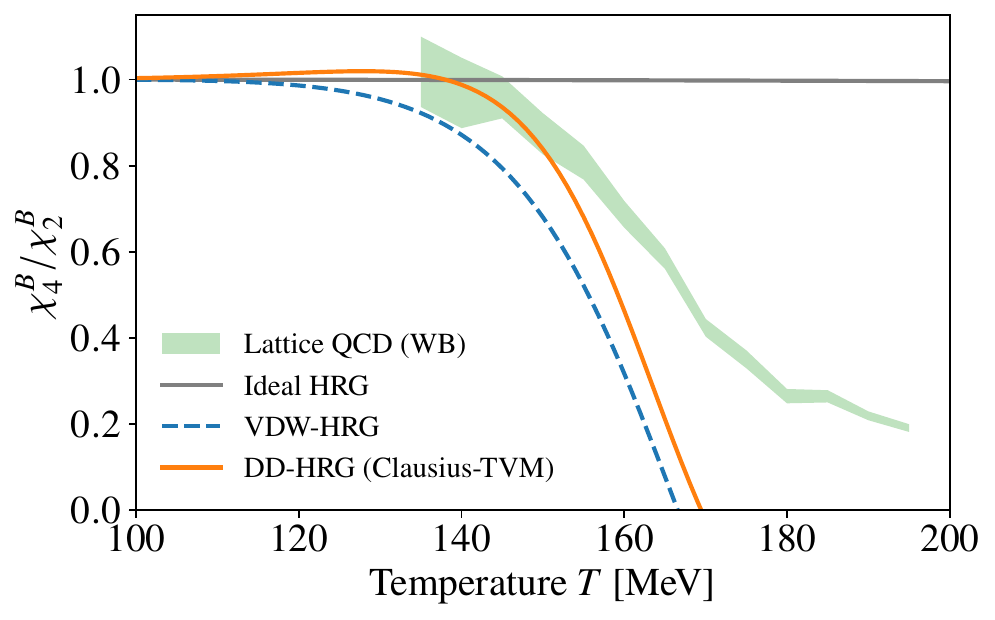}
\caption{Scaled pressure $P/T^4$ (left) and baryon number susceptibility
ratio $\chi_4^B/\chi_2^B$ (right) at $\mu_B = 0$ as functions of
temperature in the ideal HRG, VDW-HRG~\cite{Vovchenko:2016rkn}, and
DD-HRG (Clausius--TVM, $K_0 = 320$~MeV) models, compared with lattice-QCD
data of the Wuppertal--Budapest
collaboration~\cite{Borsanyi:2013bia,Borsanyi:2018grb}.}
\label{fig:eos}
\end{figure}

\section{Finite temperature at $\boldsymbol{\mu_B = 0}$}
\label{sec:finiteT}

A non-trivial check is that the parameters fixed by cold nuclear matter
also govern finite-temperature thermodynamics, where the HRG confronts
lattice QCD. Figure~\ref{fig:eos} shows the pressure and the baryon
susceptibility ratio $\chi_4^B/\chi_2^B$ at $\mu_B = 0$, with
interactions applied to baryon-baryon and antibaryon-antibaryon pairs.
The DD-HRG pressure agrees with the lattice data across the crossover
region, while the suppression of $\chi_4^B/\chi_2^B$ is milder than in
the VDW-HRG and closer to the lattice result.

\needspace{15\baselineskip}%
\begin{wrapfigure}{r}{0.38\linewidth}
\centering
\vspace{-0.8\baselineskip}
\includegraphics[width=0.97\linewidth]{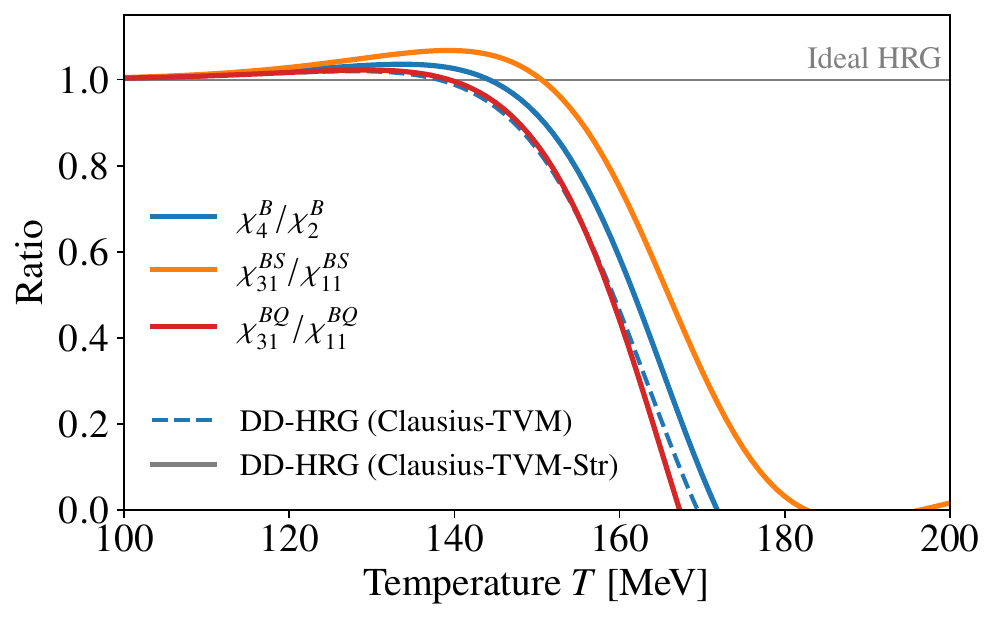}
\caption{Susceptibility ratios at $\mu_B = 0$ in the DD-HRG
($K_0 = 320$~MeV) with strangeness-dependent excluded volume,
Eq.~(\ref{eq:bij}), for $b_{BS} = 0.3$~fm$^3$ (solid lines), and without
strangeness dependence where all three ratios coincide (dashed line).}
\label{fig:chi}
\end{wrapfigure}
Ratios of fourth-to-second order conserved-charge susceptibilities probe
the flavor dependence of baryonic interactions: for interactions
independent of the baryon quantum numbers, all three ratios
$\chi_{31}^{BS}/\chi_{11}^{BS}$, $\chi_4^{B}/\chi_2^{B}$, and
$\chi_{31}^{BQ}/\chi_{11}^{BQ}$ coincide~\cite{Karthein:2021cmb}, whereas
lattice QCD suggests a definite ordering,
$\chi_{31}^{BS}/\chi_{11}^{BS} > \chi_4^{B}/\chi_2^{B} >
\chi_{31}^{BQ}/\chi_{11}^{BQ}$~\cite{Karthein:2021cmb,Borsanyi:2018grb}.
To explore this further, we add strangeness dependence to the excluded-volume matrix, parametrized as,
\begin{equation}
b_{ij} = \bar b + \delta b_I\, I_{3,i}\, I_{3,j}
         + b_{BS}\big(B_i S_j + S_i B_j\big),
\label{eq:bij}
\end{equation}
where $I_3$ is the isospin projection, $B_i$ and $S_i$ are the baryon number and strangeness of hadron species $i$, and $\bar b$ and $\delta b_I$ encode the isospin-averaged and isospin-dependent parts fixed by the nuclear matter.
A positive $b_{BS} = 0.3$~fm$^3$
implies weaker EV repulsion for hyperons ($S_i<0$) and reproduces the
lattice ordering, as shown in Fig.~\ref{fig:chi},
consistent with the earlier EV-HRG analysis of Ref. [10], 
which likewise favored reduced repulsion in the strange-baryon sector.
% that favored reduced repulsion in the
% strange baryon sector.

\section{Summary and outlook}
\label{sec:summary}

We formulated a density-dependent van der Waals HRG model allowing for an
arbitrary mean field and a generalized excluded-volume mechanism,
implemented in the open-source \texttt{Thermal-FIST} v1.6.1. 
With density- and isospin-dependent parameters constrained by nuclear-matter properties,
the model provides a description of neutron-star matter supporting
$\gtrsim 2\,M_\odot$ stars, while the same EoS improves the agreement
with lattice QCD thermodynamics at $\mu_B = 0$, where susceptibility
ratios suggest weaker repulsion for strange baryons.
Future work will constrain the flavor dependence of interactions through Bayesian analyses of combined heavy-ion, astrophysical, and lattice constraints.

\vspace{0.1\baselineskip}
\noindent\textbf{Acknowledgements.}
This work is supported by the U.S. Department of Energy, Office of Science, Office of Nuclear Physics, Early Career Research Program under Award Number DE-SC0026065.

% \begin{multicols}{2}

% \end{multicols}

\end{document}